\documentclass[apj]{emulateapj}  
\usepackage{apjfonts}
\usepackage{mathptmx}
\usepackage{psfig}
\newcommand {\lsim}{\mbox{$\:\stackrel{<}{_{\sim}}\:$} }

\newcommand {\etal}{{et\thinspace al.} }
\newcommand {\parn}{\par\noindent}

\def\li#1{\hbox{$^{#1}{\rm Li}$}}

\def\6Li{\mbox{$^6$Li}}
\def\7Li{\mbox{$^7$Li}}

\def\zstar{\mbox{$z^\star$}}
\def\Eps{\mbox{$E_s^{'}$}}

\def\Kap{\mbox{${\rm K}_{\alpha{\rm p}}$}}
\def\beq{\begin{equation}}
\def\eeq{\end{equation}}

\def \Eq#1{{Eq.~\ref{e:#1}}}     
\def \EQN#1{\label{e:#1}}        
\def \Fig#1{{Fig.~\ref{f:#1}}}   

\def \ion#1#2{#1~{\sc #2}}
\def \unit#1#2{\mbox{#1$^{#2}$}}
\def \d#1{{\rm d} #1}

\shorttitle{}
\shortauthors{}

\begin{document}

\journalinfo{
\parbox{1.5in}{
astro-ph/0605633 \\
UMN--TH--2505/06 \\
FTPI--MINN--06/18 \\
May 2006}
}

\title{Population III Generated Cosmic Rays and the Production of ${}^6$Li.}

\author{Emmanuel Rollinde\altaffilmark{1}, Elisabeth Vangioni\altaffilmark{1}, Keith A. Olive\altaffilmark{2}}
\altaffiltext{1}{Institut d'Astrophysique de Paris, UMR7095 CNRS, Universite Pierre \& Marie Curie, 98 bis bd Arago, 75014 Paris, France, rollinde@iap.fr, vangioni@iap.fr}
\altaffiltext{2}{William I. Fine Theoretical Physics Institute, School of Physics and Astronomy, University of Minnesota, Minneapolis, MN 55455, USA, olive@physics.umn.edu}

\begin{abstract}
\parn 
 We calculate the evolution of \6Li\  generated from cosmic rays produced
 by an early population of massive stars.
 The computation is performed in the framework of 
hierarchical structure formation and is based
on cosmic star formation histories constrained to reproduce the observed star formation
rate at redshift $z \la 6$, the observed chemical abundances in damped Lyman alpha absorbers
and in the intergalactic medium, and to allow for an early reionization of the Universe
 at $z\sim 11$ by Pop III stars as indicated by the third year results released by WMAP.
We show that the  pregalactic production of the \6Li\
 isotope in the IGM via these Pop III stars can account for the \li6
  plateau observed in metal poor halo stars
without  additional over-production of \li7.
Our results depend on the efficiency of cosmic rays to propagate out of 
 minihalos and the fraction of supernovae energy
deposited in cosmic rays.  We also compute the cosmic ray heating of the IGM gas.
In general, we find somewhat high temperatures (of order $10^5$ K) implying
that the cosmic rays production of \6Li may be required to be confined to the so-called
warm-hot IGM.  
 
\end{abstract}

\keywords{ Cosmology - Cosmic rays - Big Bang Nucleosynthesis -  Stars: abundances}

\section{Introduction}
\label{s:introduction}

The Lithium observed in low metallicity environments such as
the atmospheres of halo stars offers a unique probe into two very
distinct mechanisms of nucleosynthesis: the big bang and cosmic rays.
Big bang nucleosynthesis (BBN) produces predominantly \7Li while cosmic
ray nucleosynthesis (CRN) produces roughly equal numbers of the 
\6Li and \7Li isotopes. The standard lore tells us that the bulk of 
the Pop~II \7Li abundance is produced by BBN with additional
contributions at the 10\% level being supplied by galactic CRN (or GCRN).
BBN lays down the primordial abundance which dominates the Spite plateau 
\citep{spites} while GCRN supplies a metallicity dependent supplement
which would result in a small slope in Li vs. Fe regressions \citep{rbofn}.
\6Li on the other hand, produced only in GCRN would be expected to 
show a strong (log-linear) relation with Fe, with an abundance which is
a fraction of the \7Li abundance.  Indeed the first observations of \6Li
at [Fe/H] $\simeq -2$ \citep{sln1,ht1,ht2,sln2,cetal,Nissen00}
were at the expected level of \6Li/\7Li $\simeq 0.05$ \citep{sfosw,Fields99,Vangioni99}. 

This simple picture has been shaken by recent observations leading to 
two distinct Li problems.
WMAP \citep{Spergel,Spergel2} has determined the baryon density of the Universe
to high accuracy, $\Omega_B h^2 = 0.0224 \pm 0.0009$, corresponding to 
a baryon-to-photon ratio, $\eta = (6.14 \pm 0.25) \times 10^{-10}$.
At this value of $\eta$, the BBN predicted value of \7Li is 
\7Li/H = $4.27^{+1.02}_{-0.83}\times 10^{-10}$ \citep{cfo,cfo3,Cyburt04}, 
\7Li/H = $4.9^{+1.4}_{-1.2}\times10^{-10}$ \citep{Cuoco}, or
 \7Li/H = $4.15^{+0.49}_{-0.45}\times10^{-10}$ \citep{Coc}.
 These values are all significantly larger than most determinations
 of the \7Li abundance in Pop II stars which are in the range $1 - 2 \times 10^{-10}$
 \citep[see e.g.][]{spites,rbofn} and even larger than a recent determination
 by \citet{mr} based on a higher temperature scale. The second Li problem
 originates from recent observations of \6Li \citep{Asplund2} which indicate 
 a value of [\6Li] $= \log{\6Li/{\rm H}} +12 = 0.8$ which is {\em independent} of
 metallicity in sharp contrast to what is expected from GCRN models.
 In effect, we are faced with explaining a \6Li plateau at
 a level of about 1000 times that expected from BBN \citep{tsof,Vangioni99}.
 Here, we will concentrate on the latter of the two Li problems. 

Different scenarios have been
discussed to explain the abundance of \6Li\ in metal-poor
halo stars (MPHS). \citet{Suzuki} discussed  
the possibility of cosmic rays produced in shocks during
the  formation of the Galaxy, which was
consistent with \6Li\ data available at that time.
\citet{Jedamzik,Jedamzik1,Jedamzik2}, \citet{kawa}, \citet{Jedamzik3}, \citet{grant}, and \citet{posp}
consider
the decay of relic particles, during  the epoch of the big bang nucleosynthesis,
that can yield to a large primordial abundance of \6Li.
\citet{Fields} have  studied in detail the lithium
production in connection to gamma rays, using a formalism similar
to ours.

Previously, we computed the evolution of  the \6Li abundance produced by 
an initial burst of cosmological cosmic rays (CCRs) \citep{RVOI}[hereafter RVOI]. We found that
 the  pregalactic production of the \6Li\
 isotope  can account for the \6Li
  plateau observed in metal poor halo stars
without  additional over-production of \7Li.
The derived relation between
the amplitude of the CCR energy spectra and the redshift of 
the initial CCR production put constraints on the physics
and history of the objects, such 
as Pop III stars, responsible for these early cosmic rays. 
Due to the subsequent evolution of \6Li\ in the Galaxy through GCRN,
we argued that halo stars with metallicities between [Fe/H] = -2 and -1,
must be somewhat depleted in \6Li.

Here, we will employ detailed models of cosmic chemical evolution to derive
the total CCR energy as a function of redshift.  
In particular, we will make use of models discussed in \citet{Daigneal05}
which were constructed to account for the observed star formation rate (up to $z \sim 6$),
the observed SN rates (up to $z \sim 1.5$), early reionization of the intergalactic medium (IGM)
(at $z \sim 11$), and reasonable chemical abundances in 
the interstellar medium (ISM) and IGM of proto-galaxies.

These models allow us to determine the total energy injected in cosmic rays
as a function of redshift. 
Once a spectral shape for the source function is given,  we must account for the propagation of CCRs. 
Furthermore, we must distinguish between diffusion in the ISM and propagation in
an expanding IGM.
We first consider the resulting nucleosynthesis if most of the high energy particles are ejected from
the minihalos where star formation begins and escape into the IGM. 
This will lead to a physical picture similar to the assumptions made in RVOI. 
In this context, we will see that cosmic-ray heating of the IGM may place important constraints
on the scenario. Depending on the fraction of SN energy deposited in CRs and the efficiency 
of shock acceleration of CRs leading to their escape from structures, it may be necessary
to consider in addition to the IGM production of \6Li, the in situ production when 
CRs are confined to structures. 

Our paper is organized as follows: In the next section, we will
briefly describe the cosmic chemical evolution models of \citet{Daigneal04,Daigneal05}.
In section 3, we describe the general features 
of cosmological cosmic rays related to the production of Lithium in the IGM.
There we also present our detailed calculations of the \6Li abundance  based on the Pop III
production of CCRs in the IGM, at high redshift.  In section 4, we comment on the in situ production of \6Li in Pop III minihalos.
We summarize the status of the CCR origin of \6Li in section 5.

\section{SN history and Cosmological Cosmic Rays}
\label{s:history}

The cosmic star formation histories considered are based on the detailed models of 
chemical evolution derived in \citet{Daigneal05}. The models are described by a bimodal 
birthrate function of the form
\beq
B(m,t,Z) = \phi_1(m)\psi_1(t) + \phi_2(m)\psi_2(Z)
\eeq
where $\phi_{1(2)}$ is the initial mass function (IMF) of the normal (massive) component of star formation and $\psi_{1(2)}$ is the respective star formation rate (SFR). $Z$ is the metallicity. The normal mode contains stars with mass between 0.1 M$_{\odot}$ and 100 M$_{\odot}$ and has a SFR which peaks at  $z \approx 3$. The massive component dominates at high redshift. The IMF of both modes is taken to be a power law with a near Salpeter slope
so that,
\beq 
\phi_i (m) \propto m^{-(1+x) }
\label{IMF}
\eeq
 with $x=1.3$.
 Each IMF is normalized independently by
 \begin{equation}
\int_{m_\mathrm{inf}}^{m_\mathrm{sup}} dm\ m \phi_i(m)=1\ ,
\label{norm}
\end{equation}
differing only in the specific mass range of each model. 
Both the normal and massive components contribute to the chemical enrichment of galaxy forming structures and the IGM, though the normal mode is not sufficient for accounting for the early reionization of the IGM \citep{Daigneal04}.

Here, we restrict our attention to the best fit hierarchical model in \citet{Daigneal05} in which the minimum 
mass for star formation is $10^7$ M$_\odot$.  The distribution of structure masses is
based on the Press-Schechter formalism \citep{ps,jen}.
The normal mode SFR is given by
\beq
\psi_1(t) = \nu_{1} M_{struct} \exp{(-t/\tau_{1})}\ ,
\eeq
where  $M_{struct}$ is the mass of the structure (which includes dark matter), $\tau_{1} = 2.8$ Gyr is a characteristic timescale and $\nu_{1}= 0.2$ Gyr$^{-1}$ 
governs the efficiency of the star formation. In contrast, the massive mode SFR is defined by
\begin{equation}
\psi_2(t) = \nu_{2} M_{\rm ISM} \exp{\left(- Z_{\rm IGM} / Z_\mathrm{crit}\right)}\ ,
\end{equation}
where $M_{\rm ISM}$ is the mass of the baryonic gas in the structure, 
with $\nu_{2}$ maximized to achieve early reionization without the overproduction of metals or 
the over-consumption of gas. We adopt $Z_\mathrm{crit}/Z_\mathrm{\odot}=10^{-4}$ .

\begin{figure}
\begin{center}
\resizebox{\linewidth}{!}{\includegraphics{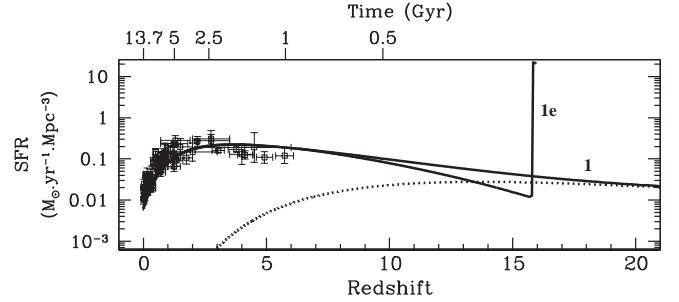}}
\end{center}
\caption{The star formation rate for Model 1, and the rapid burst Model 1e.
The dotted curve shows the SFR of the massive mode of Model 1. Data come from \cite{hopkins04}.}
\label{fig:extremModel1SFRFracBZ}
\end{figure}

We also restrict our attention to Model 1 of \citet{Daigneal05} to describe the massive mode.
In Model 1,  the IMF is defined for stars with masses, 40 M$_{\odot} \le m \le 100$ M$_{\odot}$. 
All of these stars die in core collapse supernovae leaving a black hole remnant.
The coefficient of star formation, $\nu_2$
is 80 Gyr$^{-1}$.
Star formation begins at very high redshift ($z \simeq 30$) but peaks 
at a redshift $z \simeq 12$.  Note that the absolute value of the SFR
depends not only on $\nu_2$, but also on the efficiency of outflow.  See \citet{Daigneal05} for details.
We also consider an
example of a model (with an IMF as in Model 1) in which the massive mode
occurs as a rapid burst at $z = 16$ designated as Model 1e.
Thus, most Pop III SN occur within a very short period of time at this redshift. 
This model is therefore 
similar to the SFR assumed in RVOI.
In Figure~\ref{fig:extremModel1SFRFracBZ}, 
we show the SFR, for Models 1 and 1e (including the normal mode).
Also shown by the dotted curve is the SFR for the massive mode alone in Model 1.

We do not here consider models with very massive stars in Pop III.
In \citet{Daigneal05}, two very massive mode models were considered
in addition to Model 1 described above.  In Model 2a, the massive mode consists of
stars with masses in the range 140-260 M$_\odot$, and in Model 2b, the range used was
270-500 M$_\odot$.  In Model 2b, there is significant metal enrichment \citep{ww2002}
which results in a diminished SFR and hence fewer cosmic rays available for
\6Li production.  In Model 2b, one expects total collapse and little or no production of 
cosmic rays.

The rate of core collapse supernovae (SNR)
can be calculated directly in terms of the IMF and SFR
\beq
SNR = \int_{max(8M_{\odot},m_{min(t)})}^{m_{sup}} dm \phi(m) \psi(t-\tau(m)),
\label{SNR}
\eeq
where $m_{min(t)}$ is the minimum mass of a star with lifetime, $\tau$, less than $t$. 
When a star undergoes core collapse, the mass of the remnant is determined by the mass of the progenitor.  We assume that all stars of mass $ m \ga 8$ M$_{\odot} $ will die as supernovae. For stars of mass 8 M$_{\odot} < m < 30$ M$_{\odot} $, the remnant after core collapse will be a neutron star of $ m \approx 1.5$~M$_{\odot} $. Stars with 30 M$_{\odot}<m<100$ M$_{\odot} $ become black holes with mass approximately that of the star's helium core before collapse \citep{stardeath}. We take the mass of the Helium core to be 
\beq
M_{He}=\frac{13}{24} \cdot (m-20\,M_{\odot}) 
\eeq
for a star with main sequence mass $m$~\citep{ww2002}. The supernova rate ultimately determines
the metal enrichment of the ISM and when coupled with the model of outflow also determines
the metal enrichment of the IGM.

The energy emitted in each core collapse, $E_{cc}$ corresponds to the change in gravitational energy, 99\% of which is emitted as neutrinos. In the cases where collapse results in a neutron star, $E_{cc}=3 \times 10^{53}\,$ergs. For stars that collapse to black holes, $E_{cc}$ is proportional to the mass of the black hole. For masses less than 100 M$_\odot$, we take $E_{cc} =  0.3M_{He}$ . 
We will parametrize the energy injected in cosmic rays per supernova
as
\beq
{\cal E}_{\rm CR}(m) = {\epsilon  E_{cc}(m) \over 100}\,,
\eeq
where $\epsilon$ is the fraction of energy in the remaining 1\% (i.e. energy not in neutrinos) 
deposited into cosmic rays.
Given the IMF described above, the massive mode is dominated by 40 M$_\odot$ stars for which the total
energy per SN in CRs is ${\cal E}_{III}=10^{52.8} \epsilon_{III}$ ergs.  
In contrast, the normal mode associated with Pop II, is dominated by lower mass stars
for which the energy per SN in CRs is ${\cal E}_{II}=10^{51.5} \epsilon_{II}$ ergs.  
While it is quite plausible that $\epsilon_{II}$ and $\epsilon_{III}$ differ (indeed we
would expect  $\epsilon_{II} < \epsilon_{III}$), we will for simplicity
assume  $\epsilon_{II} = \epsilon_{III} = \epsilon = 0.01 - 0.30$ as a broad and 
conservative range.

\begin{figure}[!h]
\unitlength=1cm
\centerline{\psfig{width=\linewidth,figure=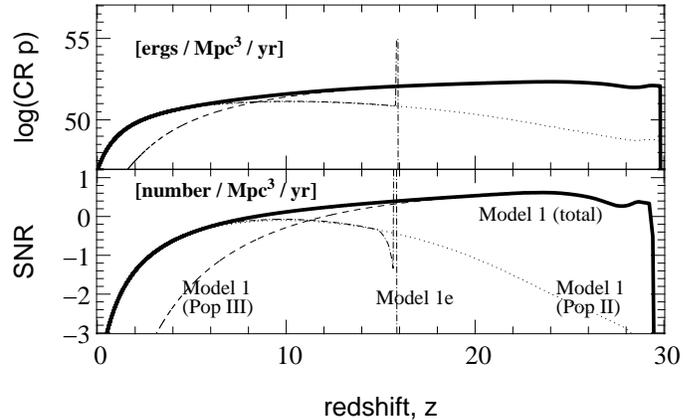}}
\caption{CR history predicted by \citet{Daigneal05}.
The SNR (lower panel) and energy density in cosmic rays (upper panel)
are shown in the case of  Model 1 for Pop III (dashed), Pop II (dotted) and 
all SN (solid) and in the case of the model 1e (dot-dashed). 
}
\label{f:daigne}
\end{figure}

The SNR derived from Eq. \ref{SNR} is shown for both Models 1 and 1e in 
\Fig{daigne} (lower panel).  In the upper panel of \Fig{daigne}, we show the 
energy density in CRs injected per year. 
The CR production rate in Model 1e (shown by the dot-dashed curves) is similar to that
 Model 1 (shown by the solid curves) below a redshift of about 10, as would
 be expected from the SFRs shown in Fig.~\ref{fig:extremModel1SFRFracBZ}.
Dashed lines show the rates of 
CRs generated by massive Pop III SNe alone, while dotted lines correspond to Pop II SNe ejection.
The energy density in cosmic rays is dominated at large redshift
by Pop III SNe 
 due to the corresponding IMFs and mass range
 associated with the two modes and the dependence of $E_{cc}$ on the progenitor mass.

\begin{figure}[!h]
\unitlength=1cm
\begin{picture}(12,4)
\centerline{\psfig{width=8.6cm,figure=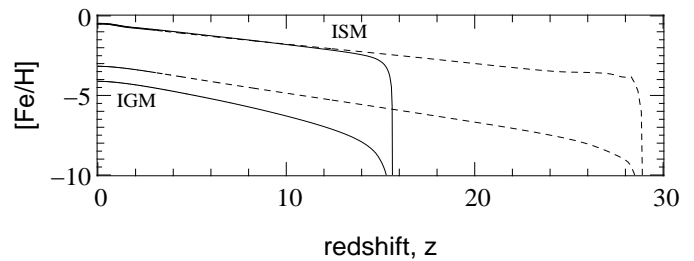}}
\end{picture}
\caption{Evolution of metallicity in ISM and IGM versus redshift in Model 1 (dashed curves) and in Model
1e (solid curves).}
\label{f:daigneFeH}
\end{figure}

The metallicity evolution in both the IGM and the ISM has been derived by \citet{Daigneal05} and is shown 
in  \Fig{daigneFeH}. For Model 1e, the metallicity in the ISM rises very quickly to 
[Fe/H] $\sim -2.5$, whereas in Model 1, the initial enrichment occurs 
rapidly only to [Fe/H] $\sim -4$.  The IGM abundance are several thousand times smaller.
Both models have the same metallicity as a function of redshift
  below $z\sim 15$ in the ISM.

\section{Production of Lithium in the IGM}
\label{s:IGM}

In RVOI,
 we considered a single burst of CRs whose total energy 
 was fixed in an ad hoc way so as to reproduce the observed \6Li\ abundance. 
 We now relate the production of CRs to the detailed model of cosmic
 chemical evolution in which the SN history is completely
 determined by the SFR and IMF of the model and are
 constrained to reionize the IGM at a redshift $z \simeq 11$, match the observed SFR 
 at $z < 6$, as well as chemical abundances at $z <3$.  As a consequence,
 the energy density in cosmic rays is determined by the model
 and we can derive the abundance of Lithium produced in the IGM
 as described below.

\subsection{Formalism of Cosmological Cosmic Rays}
\label{s:formalism}

 \subsubsection{Ejection of CCR into the IGM}
 
The total kinetic energy given initially to CRs by SN is 
\begin{equation}
{\cal E}_{\rm SN}(z)\,= (1+z)^3 \int_{max(8M_{\odot},m_{min(t)})}^{m_{sup}} dm \phi(m) \psi(t-\tau(m)) {\cal E}_{\rm CR}(m).
\label{crenergy}
\end{equation}
where it is understood that the appropriate IMF is used for computing CR energy density due to Pop~II or Pop~III SN.
For example, using Eq. \ref{crenergy} we can estimate from Fig. \ref{fig:extremModel1SFRFracBZ},
the Pop~III contribution to the CR energy density.
In Model 1, if we take $\psi \simeq 0.02$ M$_\odot$ yr$^{-1}$ Mpc$^{-3}$
from $z = 10 - 30$, and approximate $\phi$ by a delta function at $m = 40$ M$_\odot$, 
we obtain ${\cal E}_{\rm SN}(10) \simeq 7.8 \times 10^{-13} \epsilon$ ergs/cm$^3$. 
Note that the result of the full calculation of Eq. \ref{crenergy} is a factor of
about 5 larger than this, because the contributions of more massive
stars (more massive than 40 M$_\odot$) can not be neglected.
This corresponds to ${\cal E}_{\rm SN}\simeq 3.6 \times 10^{-12} \epsilon$ ergs/cm$^3$.
For Model 1e, we see that the burst is very intense, with a SFR which
reaches 20 M$_\odot$ yr$^{-1}$ Mpc$^{-3}$, though the duration is only
about $3 \times 10^6$ yr.  In this case, we approximate that 
${\cal E}_{\rm SN}(16) \simeq 1.6 \times 10^{-11} \epsilon$ ergs/cm$^3$.
The full calculation yields ${\cal E}_{\rm SN} \simeq 2.5 \times 10^{-11} \epsilon$ ergs/cm$^3$.

\bigskip

\subsubsection{Source Spectrum}

We next briefly describe our treatment of cosmological cosmic rays, their transport and
the mechanism for the production of \6Li, in the case of one burst of CRs at 
a given redshift $z_o$.
Our formalism is directly derived from 
the work of  \citet{Montmerle}, hereafter M77. 
We begin by defining the source function for our CR distribution.
The source function  $Q(E,z)$ is defined as
 a power law in momentum,
\begin{equation}
Q(E,z)\ = \ C(z_o) \frac{(E+E_0)}{\left(E(E+2\,E_0)\right)^{(\gamma+1)/2}} \delta(z-z_o) [\unit{GeV}{-1}\unit{cm}{-3}\unit{s}{-1}]\,,
\label{source}
\end{equation}
where $C(z)$ is fixed by our normalization of the source function 
to the total energy from both Pop~II and Pop~III SNe  using 
\begin{equation}
{\cal E}_{\rm SN}(z) ={\cal E}_{\rm SN,II}(z) + {\cal E}_{\rm SN,III}(z) =  \int_{E_{\rm min}}^{E_{\rm max}} E\,Q(E,z)\,\d{E}\ 
\end{equation}
We take $E_{\rm min}=0.01$ MeV and $E_{\rm max}=10^6$ GeV \footnote{Note that in  RVOI we used $E_{\rm min}$ =  10 MeV. When the integral diverges, i.e. when the power law $\gamma$ is larger than 2, this can affect the result by a factor as large as  2. The cut-off of  0.01 MeV is related to a minimum kinetic energy for CRs to escape the star itself.}. We set $\gamma=3$ in this section.

The efficiency of the SN to eject CRs outside the structure 
and into the IGM
depends on many different physical parameters. Roughly, low energy particles will
lose all their energy inside the structure and in this section, we take a simplified 
approach where the CRs spectrum is simply cut at a given energy, $E_{\rm cut}$
and decreased by a constant factor $\epsilon_{\rm shock}$.
\begin{equation}
Q_{\rm IGM}(E) = 
\left\{
\begin{array}{ll}
0 & {\rm if\ } E<E_{\rm cut} \\
\epsilon_{\rm shock}\ Q(E) & {\rm if\ } E\geq E_{\rm cut} 
\end{array}
\right.
\end{equation}
In \cite{Daigneal05}, the efficiency of the baryon outflow rate coming from the structures
is dependent on the redshift. It accounts for the increasing escape velocity of 
the structure as the galaxy assembly is in progress. In fact, 
in \citet{Daigneal04} two sources of outflow were considered : a global
 outflow powered by stellar explosions (galactic winds) and an outflow corresponding to 
 stellar supernova ejecta that are pushed directly out of the structures as chimneys. 
However, velocities in ISM gas are of the order of a few 100 km/s  while 
 CRs are mostly relativistic. Thus, the CR ejection processes 
considered here should be independent of the overall outflow of gas and heavy elements. 
As we will see, the production of Lithium is proportional to the energy available, and thus to $\epsilon_{\rm shock}$.  
   Most of the CR production will occur at large redshift, when Pop III are dominant (especially
   in the case of the Model 1e).
 For simplicity, we will assume that all CRs are ejected from structures and adopt a constant
value for $\epsilon_{\rm shock} = 1$.  We comment on the possibility of $\epsilon_{\rm shock} < 1$
in \S 4.

\subsubsection{Production of Lithium in the IGM}

If $N_i(E,z)$ is the comoving number density per (GeV/n) of a given
species at a given time or redshift, and energy, we define $N_{i,{\rm H}}(E,z)\, \equiv
\, N(E,z)/n_{\rm H}(z)$, the abundance by number with respect
to the ambient gaseous hydrogen (in units  of (Gev/n)$^{-1}$).
The evolution  of $N_{i,{\rm H}}$  is defined through the transport
function 
\begin{equation}
\frac{\partial N_{i,{\rm H}}}{\partial t} + \frac{\partial}
{\partial E}(bN_{i,{\rm H}}) + \frac{N_{i,{\rm H}}}{T_{\rm D}} = Q_{i,{\rm H}}\,.
\label{e:transport}
\end{equation}
where ${T_{\rm D}}$ is the lifetime against destruction and 
$b$ describes the  energy losses due to expansion or ionization
processes ((Gev/n)\,s$^{-1}$).
The energy and time dependencies can be separated 
 as $b(E,z)=-B(E)f(z)$.
We can distinguish two cases depending on 
whether losses are dominated by expansion or by ionization. 
The general form for the redshift dependence, when 
expansion dominates
is $f_{\rm E}(z)=(1+z)^{-1}|\d{z}/\d{t}|\,H_0^{-1}$
\citep[e.g.][]{Wick}. Other contributions to 
$B$ or $f$, do not depend on the assumed cosmology
and are given explicitly in M77.

In M77,  two important quantities, $\zstar(E,E',z)$ and 
$\Eps(E,z)$ are used in this formalism. Given 
a particle ($\alpha$ or lithium) with an energy $E$ 
at a redshift $z$, $\zstar(E,E',z)$
corresponds to the redshift at which this particle
had an energy $E'$. $\Eps(E,z)$ is the initial energy required 
if this particle was produced  at the redshift of the burst,
$z_s$. In particular, $\zstar(E,\Eps,z)=z_s$.
The equation that defines $\zstar$ (Eq. A5, M77) is
$\partial{\zstar}/\partial{E} =  - \left[B(E)f(z)\,|\d{z}/\d{t}|\right]^{-1}\, \left(\partial{\zstar}/\partial{z}\right)$.

The evolution of the CCR $\alpha$ energy  
spectrum is derived, using Eq. A8 of M77 
\begin{equation}
\Phi_{\alpha, {\rm H}}(E,z) = {\phi_\alpha(E)\over \ n_{\rm H}^0}\frac{\beta}{\beta'}\frac{\phi_\alpha(\Eps)}{\phi_\alpha(E)}\left|\frac{\d{z}}{\d{t}}\right|_{z_s}\frac{\exp{(-\xi)}}{|b(E,z_s)|}\,\frac{1}{\left|\partial\zstar/\partial E'\right|_{{E_s^\prime}}}
\EQN{phialphaevol}
\end{equation}
where $\Phi_{\alpha, {\rm H}}(E,z)\equiv \Phi_\alpha(E,z)/n_{\rm H}(z)$
 is the flux of $\alpha$'s per comoving volume
\begin{equation}
\Phi_{\alpha, {\rm H}}(E,z) = \beta\,N_{\alpha,{\rm H}}(E,z)
\EQN{phialphadef}
\end{equation}
and  $\beta$ ($\beta'$)
is the velocity corresponding to energy
$E$ ($\Eps$); $\xi$ accounts for the destruction term (Eq. A9,
 M77). The CR injection spectrum, $\phi_\alpha$ is proportional to the source spectrum, $Q$.

The evolution of CR flux and the production of Lithium 
can be computed step by step 
directly with the transport function (\Eq{transport}).
The abundance by number of lithium
($l$\,=\,\6Li or \7Li) of energy $E$, 
produced at a given redshift $z$,
is computed from 
\begin{eqnarray}
{\partial N_{l,\, {\rm H}}(E,z) \over \partial t} & = & \int \sigma_{\alpha\alpha\rightarrow l}(E,E')n_{\rm He}(z)\Phi_{\alpha\, ,{\rm H}}(E',z)\,\d{E'} \nonumber\\
                      & = & \sigma_l(E) K_{\alpha p} \Phi_{\alpha}(4\,E,z)\ \ [\mbox{(Gev/n)$^{-1}$\,s$^{-1}$}]\,,
\end{eqnarray}
where $\sigma_{\alpha\alpha\rightarrow l}(E,E')=\sigma_l(E)\delta(E-E'/4)$
and $K_{\alpha p}=0.08$ is 
the abundance by number of $^4$He/H.
We use cross sections based on recent 
measurements related to
the $\alpha+\alpha$ reaction and provide a new fit for the production of  \6Li\
and \7Li \citep{Mercer}.

If we now consider the contribution of each individual
burst at each redshift :
\begin{equation}
\begin{array}{lcl}
 {\rm (Li}/{\rm H)}(z) &=& {\rm Li}/{\rm H}_{\rm BBN}+\int_{z'>z}\d{z'}\int_z^{z'}  \d{z''}\nonumber\\
 && \times \int_E \, \sigma_l(E) K_{\alpha p} {\Phi_{\alpha}}(4\,E,z'')\, \d{E}\, |\d{t}/\d{z''}|\, |\d{t}/\d{z'}|\,\,, \\
\end{array}
\EQN{lh}
\end{equation}

\subsection{Results}
\label{s:igmresults}

Although the exact evolution of CR confinement is 
difficult to estimate, \cite{ensslin03} discusses the relation of the 
diffusion coefficient with a magnetic field. 
\cite{jubelgasal} propose a simple model where
this coefficient varies as the inverse of the square root of the density.
The efficiency of the diffusion will then decrease with the density (as could
be intuitively inferred). Then, at large redshift, the structures
are smaller, less dense \citep[e.g.][]{Zhao03}
and the primordial magnetic field could be expected
to confine less
than it does today, which corresponds to a large value for 
$\epsilon_{\rm shock}$. As noted above, we will assume that 
$ \epsilon_{\rm shock} = 1$, bearing in mind that this approximation
should not be valid at low redshifts ($z \la 3$). 
In effect, our results for the production of Li will depend
on the product of the CR escape efficiency, $\epsilon_{\rm shock}$,
and the efficiency for converting SN energy into CRs, $\epsilon$. 
We will return to discuss the value 
of $\epsilon_{\rm shock}$ further below.

In this work, we have also introduced the energy cut-off parameter $E_{\rm cut}$. 
Its influence is actually straightforward. If $E_{\rm cut}$ is below 10 MeV, it 
does not modify our results for the Li abundance. 
Given the shape of the Li production cross section, 
only $\alpha$ particles around 10 MeV  will produce Lithium. If
particles have a lower energy initially, they will never be available to
create Lithium. While if they are at higher energy, they will loose their
energy during their transport through the medium, and at some time reach 
the optimum energy. Thus, if the cut-off is larger than 10 MeV, there will be a delay 
until some particles loose enough energy to produce Lithium efficiently.
As shown in  RVOI,  the production of Li decreases very rapidly with time due to the 
global expansion, as a result this delay is  unfavorable to Lithium 
production. 

\begin{figure}[!h]
\unitlength=1cm
\begin{picture}(12,8)
\put(0,0){\psfig{width=\linewidth,figure=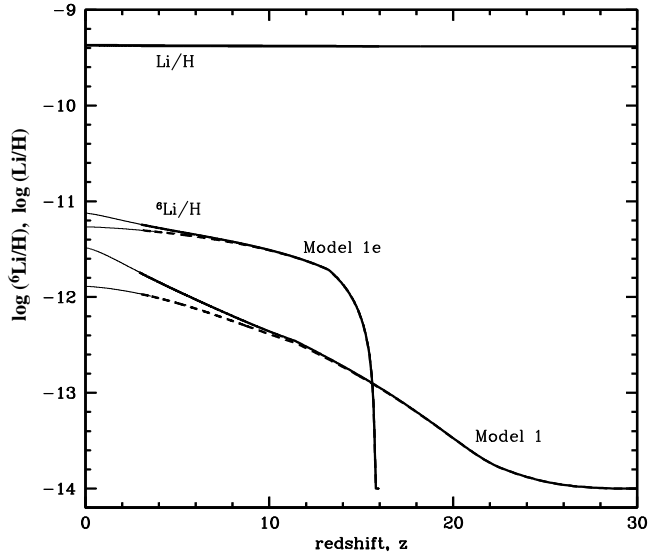}}
\end{picture}
\caption{The production of Lithium in the IGM by CCRs assuming 
$E_{\rm cut} < 10$ MeV and $\epsilon_{\rm shock}=1$ 
as a function of  redshift for both isotopes in both Models 1 and 1e. 
Here, $\epsilon$ is fixed at 0.04 so that Model 1e produces $\log$ \6Li/H = -11.2
at $z = 3$.  Since the Li abundance scales with $\epsilon$, adopting $\epsilon = 0.15$,
would increase the Model 1 abundance to -11.2 as well.  The contribution from Pop III stars alone 
to the Li
abundance is shown by the dashed curves. }
\label{f:igm2}
\end{figure}

Pop III production of CRs dominate over Pop~II at high redshift and
most of the production of \6Li\
is due to the Pop III SN. 
Model 1e is closer to our RVOI assumption of a single burst.
In \Fig{igm2}, we show the evolution of the Li abundances for $\epsilon_{\rm shock}=1$
in Model 1 and Model 1e as labeled.  
The CR efficieny, $\epsilon$  has been fixed at  $\epsilon = 0.04$ so that
in Model 1e, assuming that 
$E_{\rm cut}<10$ MeV, the total amount of Lithium
produced at $z=3$ is [\6Li] =-11.2. 
This value is perfectly consistent with the observations of \6Li in halo stars.
In Model 1e, the sum of  all energy within the burst of CRs at $z\sim 16$
corresponds to $10^{-12}$ ergs\,cm$^{-3}$  (for $\epsilon=0.04$),
and according to RVOI  ($6.3\times10^{-13}$ at $z=10$) would result in producing \6Li
at a value similar to that in the plateau.

 In Model 1, the total energy,
 integrated over the full star formation history, 
 is only about a factor of 3 less than that found in Model 1e. 
 Consequently, the production of Lithium
is reduced to   $\epsilon(\6Li)=-11.7$ at $z=3$, 
for $\epsilon = 0.04$
and thus is consistent with observations for an increased value of 
 $\epsilon = 0.15$. 
 These values are consistent with expectations that roughly 10\% of non-neutrino SN energy
 is converted to CR acceleration \citep{drury}.
 For $\epsilon = 0.15$,  the energy density in cosmic 
 rays in Model 1 is $5.4 \times 10^{-13}$ ergs\,cm$^{-3}$
 which is still slightly 
lower than the estimate in RVOI for a burst  at $z=30$ and
very similar to the one needed at $z=10$ to produce the \6Li plateau. 
Note that, as claimed in RVOI, \7Li is not overproduced by this process.

In \Fig{igm2}, we also show the Pop III contribution alone to the Li production (dashed curves) for
both Models 1 and 1e.  As one can see, Li production is largely dominated by the Pop III
contribution, though at lower redshifts, the production from the normal mode is
non-negligible. If our calculation is extended to $z=0$, we obtain a small enhancement in
the \6Li abundance as shown by the thin curves in \Fig{igm2}.  
However, as noted earlier, below $z = 3$, we expect that as the structures are 
larger and contain more baryons, CR escape will be limited resulting in a smaller
value for $\epsilon_{\rm shock}$.  In this case, we expect the in situ production of 
Li in the ISM to dominate as discussed below in \S 4. 

For $E_{\rm cut}=100$ MeV, the production
would be  decreased by one order of magnitude. 
Had we chosen a spectral index $\gamma = 2$, our results for both Models 1 and 1e
would be diminished by a factor of about 40. 

\begin{figure}[!h]
\unitlength=1cm
\begin{picture}(12,8)
\put(0,0){\psfig{width=\linewidth,figure=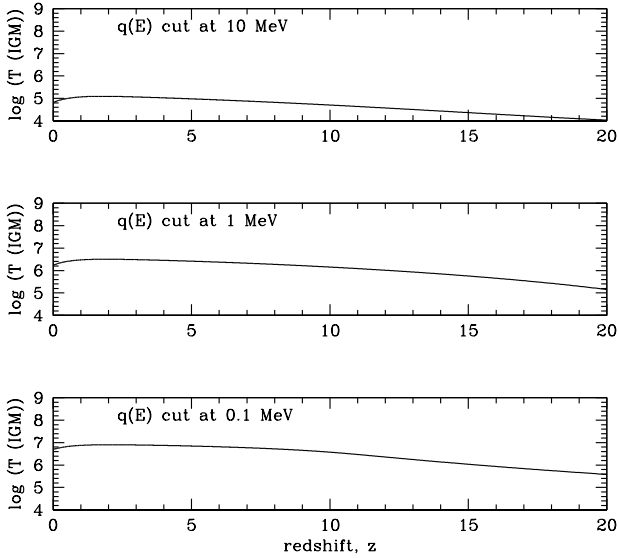}}
\end{picture}
\caption{The induced temperature by CCR heating in the IGM for Model 1
 for three choices of the CR cut-off energy, $E_{\rm cut}$ as indicated.
}
\label{f:temp}
\end{figure}

\begin{figure}[!h]
\unitlength=1cm
\begin{picture}(12,8)
\put(0,0){\psfig{width=\linewidth,figure=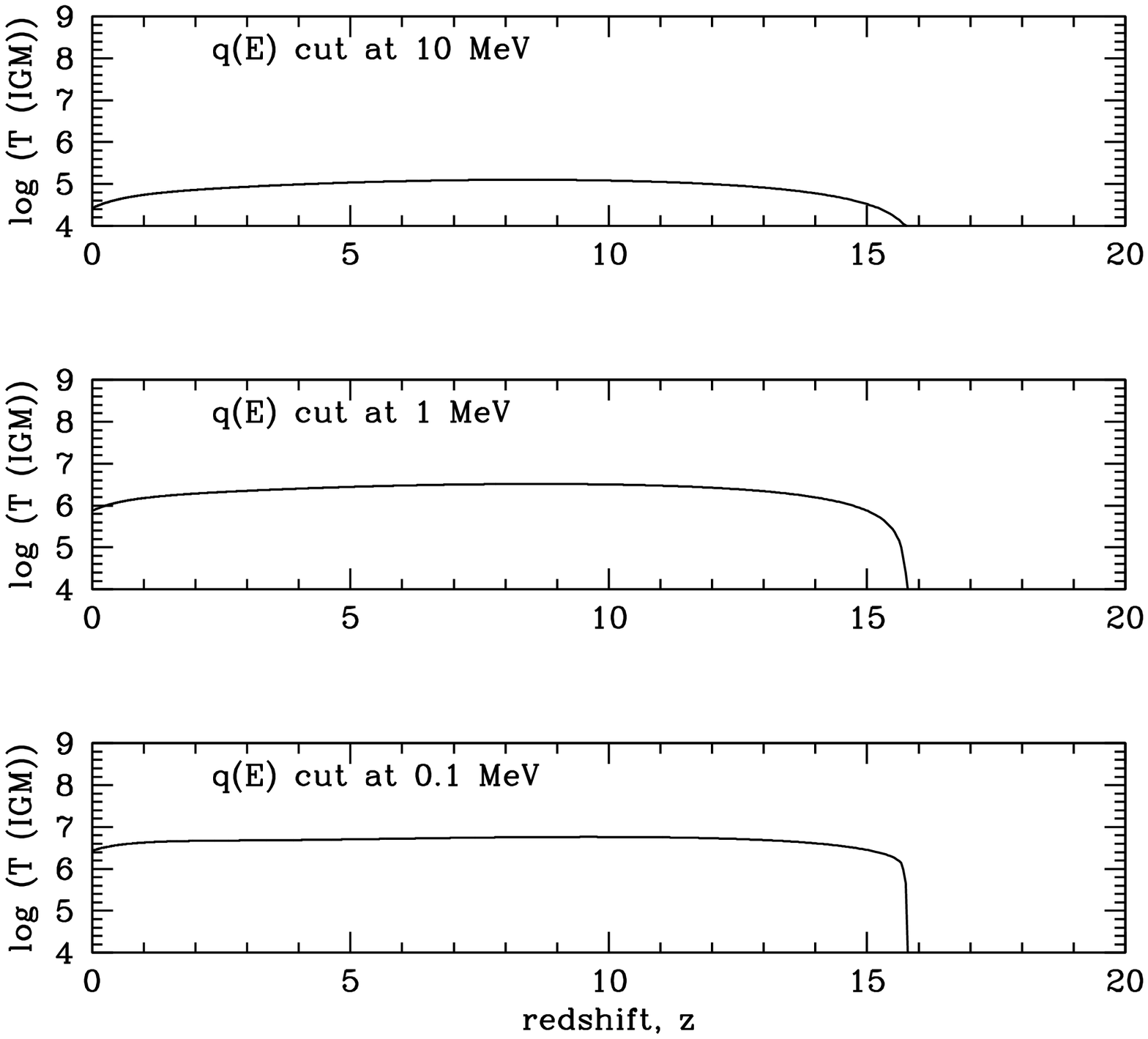}}
\end{picture}
\caption{As in \Fig{temp} for Model 1e.}
\label{f:temp2}
\end{figure}

The injection of CRs in the IGM will heat the medium. 
In fact,  CCRs were predicted to heat the IGM and thus 
avoid the problem of overcooling in  the IGM gas \citep{blanchard}.
 Following the analysis of \citet{Samui05}, we find
 that the temperature reaches  as high as $10^7$~K in Model 1
at $z\sim 5$ when $E_{\rm  cut} \la 0.1$ MeV
and $\epsilon_{\rm shock} =1.0$.
  There is a relatively
strong correlation between the induced temperature and the
CR energy cut-off as seen in \Fig{temp} where the temperature
due to CR heating is shown as a function of redshift for
three choices of $E_{\rm cut}$.  For $E_{\rm cut} = 10$ MeV, we see that
the temperature is held to the range  $10^{4.5} - 10^5$ K at $z=0$.
Because we have fixed $\epsilon = 0.04$  in Model 1e (as opposed to 0.15 in Model 1), 
we find somewhat lower temperatures for Model 1e as seen in \Fig{temp2}. In addition,
the lower temperature is partially due
to the difference in the SN history: in Model 1e, there is a sudden heating of the IGM that can cool for a longer time than in Model 1 for which heating is more progressive.
 Observations of absorption lines in quasar spectra at $z\lsim 4$
 set a conservative upper limit on the temperature of the IGM
 of $10^5$ K \cite[e.g.][]{schaye,rollinde,theuns}. 
 Thus both Models 1 and 1e, with $E_{\rm cut} = 10$ MeV and $\epsilon_{\rm shock} = 1.0$
 will not overheat the IGM.
 
The IGM temperatures shown in \Fig{temp} and \Fig{temp2} should
be viewed as an upper limit to the temperature in the IGM 
produced by CR heating.  Firstly, to produce \Fig{temp} and \Fig{temp2},
we assumed $\epsilon_{\rm shock} = 1.0$ at all times, including $z < 3$.
At low redshift, the Press-Schechter formalism
breaks down as structures are no longer representative objects.  As larger
galaxies form, the ability to eject CRs diminishes and we expect $\epsilon_{\rm shock}$
to decrease.  For $\epsilon_{\rm shock} = 0$ below $z = 3$, the temperatures
would be lower by a factor of about 2 at $z = 0$.  

Secondly, our computation 
assumes that CRs propagate into the IGM in a homogeneous way.
This is certainly not the case, as we expect
the cosmic ray density and heating to be confined to 
the warm-hot IGM (WHIM).  Correspondingly, 
the production of \6Li in the IGM may also occur 
in the WHIM embedding the structures.
In fact, the temperatures shown in \Fig{temp} are quite 
representative of the temperatures found in the WHIM
\citep{cen99,simcoe}.  The heating of cluster gas by CRs
used to produce \6Li was considered in \citet{nath}.

\subsection{Discussion}

\begin{figure}[!h]
\unitlength=1cm
\begin{picture}(12,8)
\put(0,0){\psfig{width=\linewidth,figure=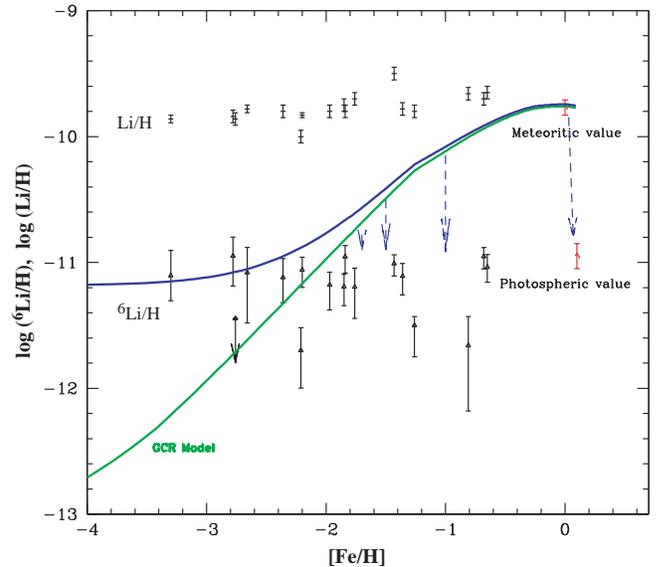}}
\end{picture}
\caption{The evolution of \6Li/H vs. [Fe/H] for 
standard galactic cosmic rays alone (shown by the solid green curve) and 
the addition of a PIE by the production of Lithium in the IGM
as in Model 1 (shown by the solid blue curve) with $\epsilon = 0.15$.
Model 1e would yield a nearly identical result with $\epsilon = 0.04$.
\6Li data come from \cite{Asplund2} with the exception of 
the lowest metallicity star which comes from \cite{inoue}.}
\label{f:igmFeH} 
\end{figure}

A key question pertains to the propagation of CRs  into 
the  IGM and the degree to which the Lithium produced is accreted onto
the Galaxy.
We assumed that all CRs  with energies above $E_{\rm cut}$
escape the structures and proto-clusters. Hence we have a homogeneous
flux of cosmic rays in the IGM. More likely, in an inhomogeneous model,
our flux would be contained in the WHIM. 
This scenario allows the Lithium produced to be accreted later,
during the formation of the Galaxy. It then provides a 
prompt initial enrichment (PIE) at $z\sim 3$ 
required to explain the observed abundances.
We can add this PIE to the standard GCR production of Lithium as
in RVOI. 
In \Fig{igmFeH}, we show the resulting evolution of \6Li as a function of 
[Fe/H].  The upper curve (solid blue) corresponds to Model 1 with $\epsilon = 0.15$ and
does an excellent job of fitting
the observed \6Li abundances at low metallicity.
The corresponding curve for Model 1e is nearly identical when $\epsilon$ is taken to be 0.04.
  A standard GCR model of \6Li production without a Population III enrichment 
 is shown in Fig. \ref{f:igmFeH}  for comparison.

As in RVOI, we require some depletion of \6Li at higher metallicities.
This should not be particularly surprising since all models GCRN predict 
a linear growth of [\6Li] versus [Fe/H]. In standard GCRN, the 
\6Li abundance matched the observations only for [Fe/H] $\simeq -2$. 
This fact was often associated with an energetics problem 
concerning the production of \6Li \citep{ramaty,ramaty2,focv}.
Pop III production of \6Li can successfully account for the \6Li plateau
at low metallicity  and together with GCR production 
accounts for the present-day meteoritic value. As 
a result, we have overproduced \6Li at [Fe/H] $\sim -1$.
This could be accounted for by depletion which is expected to be
important at higher metallicities. It is also plausible that the degree 
of depletion increases with increasing metallicity \citep{piau}.

The large temperatures produced in the IGM, prohibit
the homogenous production of \6Li (if  $E_{\rm cut}<10$ MeV).
If one considers that the \6Li production processes occur in the WHIM, 
the constraint on the temperature are very much relaxed. 
Indeed, this medium is denser, located around the galaxies or in 
filaments of clusters of galaxies. Hydrodynamic   simulations \cite[e.g.][]{cen99,dave01,kangal05} 
have long since shown that 
the WHIM, heated by shock waves during structure formation,
 displays two different phases: a cold phase, at T$\lsim\ 10^5$ K
and a hot phase, at T$\sim\ 10^{5-7}$K. 
The cold one is observed through the \ion{O}{vi} absorption lines
whose width distribution constrains the temperature \cite[e.g.][]{bergeron}.
The hot one may be responsible for the soft X-ray background, observed by {\em Chandra} 
\cite[e.g.][]{nicastroal,mckernan} and its temperature can be constrained 
by the observation of broad \ion{H}{I} lines \cite[e.g. with FUSE, ][]{richter}
or of \ion{O}{vii} lines with {\em Chandra} \citep{fang}
and, in the future, with XEUS \cite[e.g.][]{viel}.
Consequently, the temperatures derived in our scenario, when taken
in the context of the WHIM, are acceptable.
In addition, in a dense region, other processes, such as H2 line cooling, 
could lower the temperature even more. Note that in \cite{soltan} hydrodynamical 
simulations indicate that a substantial fraction of baryons in the universe remains
 in a diffuse component WHIM. This component is predicted to be at T$\sim\ 10^{5-7}$K, as noted previously.

Finally,  one should bare in mind that we set $\epsilon_{\rm shock}$ equal
to a constant value (1.0).
The exact value of $\epsilon_{\rm shock}$ depends on the strength
of the magnetic field at each epoch, on the density of the ISM etc...
 It is difficult to accurately estimate the escape fraction,
which is certainly dependent on the energy of the particles 
(roughly modeled here 
by $E_{\rm cut}$). 
\citet{ensslinal} have begun an investigation into
the propagation of CRs in the ISM taking into account all physical components. To date, they are
only able to obtain the mean values of 
the physical parameters, but this is certainly a path to follow. 
Note that if the efficiency is much less than one, the induced IGM temperature is reduced, as is the
initial enrichment of \6Li.
This will be partially offset because the density in the WHIM is larger 
than the mean density of the universe (used in the above formalism), 
and it is quite plausible that CRs are trapped there
and increase the initial enrichment of Lithium.

\section{Production of Lithium in the ISM}
\label{s:ISM}

The model discussed in the previous section
was a homogeneous model in which the IGM flux of CCRs was controlled
by the SN history and in particular, our choices of 
$\epsilon_{II}$,  $\epsilon_{III}$ and as discussed above, 
$\epsilon_{\rm shock}$.
If CR propagation into the 
WHIM is efficient, i.e. $\epsilon_{\rm shock} =1$, their
interaction with the medium will produce the required amount of 
Lithium as a PIE, as discussed in the previous section.

In fact, we expect CR propagation to be limited by diffusion, particularly at lower 
redshift so that they
are concentrated in or near the structures leading to not only enhanced
\6Li production but also heating of the eventual intra-cluster medium as well as 
the warm-hot IGM.
In this section, we will outline the computation of \6Li\ production in the structures 
of the hierarchical formation scenario that end with our Galaxy.
We assume that the CR energy output of  SN is constant 
over a sufficiently long time so that we can work in the context of a leaky box model. 
We can then adopt the physical formalism developed for 
LiBeB production in our Galaxy \citep{MAR}, hereafter MAR, \citep{elisa,fos}.

In GCRN, it is common to normalize the flux
of CRs by reproducing the observed 
abundance of Beryllium at present. 
In contrast, here, we rely on the same cosmological models described above 
when we consider the cosmological evolution of CR and the \6Li abundance.
In addition, we must take into account  the evolution with redshift
of several physical parameters defined below.

Finally, the production of Lithium in the IGM as described in the previous section
acted as an effective prompt initial enrichment of 
\6Li in the IGM which by subsequent accretion and growth of structure
led to the \6Li\ plateau observed in our Galaxy.
In the case of ISM production, each structure along the hierarchical tree 
inherits the medium as modified by SN that explode in the past. This explains the relation 
between metallicity and redshift  seen in \Fig{daigneFeH}. 
Thus, when we observe Lithium in a star at a given
metallicity, we must use the Lithium abundance present when this same metallicity was
reached. 

\subsection{Formalism}

As before we compute the total rate of energy density 
put into CRs from SN, ${\cal E}_{\rm SN}$, using
Eq. \ref{crenergy}. We use  the 
same source term, $Q(E,z)$, defined in Eq. \ref{source}.
In the framework of a diffusion model
in a medium of 
density $\rho$, it is useful to consider the scaled source
$q=Q/\rho$.
The normalization of $Q$ is related to the energy density 
 inside the structures. However,
 ${\cal E}_{\rm SN}$ corresponds to the energy density provided by SN {\em
if it was uniformly distributed within the universe},
which was true when considering the diffusion into the IGM previously.
Since this energy is now confined within the structure,
 the density there must be larger by a factor  $\rho/f_b \rho_b$,
 where $\rho_b$ is the average baryon energy density in the Universe
 and $f_b$ is the fraction of baryons found in structures.
 Therefore source function is normalized using 
 \begin{equation}
\int E\,q(E,z)\, \d{E} \ =\ \frac{{\cal E_{\rm SN}}(z)}{f_b(z)\,\rho_c\,\Omega_b} .
\end{equation}
Note that all parameters that vary with redshift are 
given with the  hierarchical model provided by \citet{Daigneal05}.

By interaction with  the particles present inside the medium, 
CR particles lose their energy. The rate of energy lost
is noted $b$. We update the relations in MAR using
\citet{mannheim94}. Note that this rate depends on the ionization 
fraction of the medium, $x_e$. We use here $ x_e=0.01$ but 
we checked that one could go up to $x_e=0.1$ without modifying
our results.
It is also convenient to define
the quantity $w=b/\rho v$ where $v$ is the velocity of the CR particle.
Because of the physical properties of the medium and the presence of a  
magnetic field, CRs will be confined to some characteristic  
escape length, $\Lambda$  which in GCRN may range from 10 - 1000 g\unit{cm}{-2}.
Under these condition, the solution of the diffusion equation (MAR) for the flux $\Phi_\alpha$ is 
\begin{equation}
\begin{array}{lcl}
\Phi_\alpha(E,z) & = &  \frac{1}{w} \int \d{E'} q_{\alpha}(E,z) \exp{\left(-\frac{R(E')-R(E)}{\Lambda(z)}\right)}\  \\
        & = & \frac{{\cal E_{\rm SN}}(z)}{f_b(z)\,\rho_c\,\Omega_b}  \, \Kap\, g(\Lambda,x_e,\gamma,z)
\end{array}
\label{e:phism}
\end{equation}
The term $q_{\alpha}=\Kap\,q$ (evaluated appropriately in the ISM)
is the source of $\alpha$ particles; and 
$R(E)=\int_0^E \d{E}/w$ is the ionization range which characterizes the average amount of 
material that an $\alpha$ particle with energy $E$ can travel before ionization losses will stop it.
 $g$ is a function that corresponds to the diffusion solution.

\begin{figure}[!h]
\psfig{width=\linewidth,figure=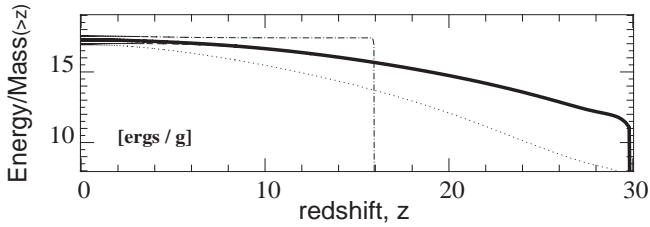}
\caption{The cumulative ratio of the energy deposited in CRs to the mass of the
structure as determined by the integral in Eq. \protect\ref{epg} (curves are as in \Fig{daigne}). }
\label{f:upper} 
\end{figure}

The rate of Lithium production of Lithium is then
\begin{equation}
\begin{array}{lcl}
\frac{\d}{\d{t}}(\frac{\rm Li}{\rm H})(z)& = & \left[\frac{\alpha}{\rm H}\right] \int \sigma\, \Phi_\alpha(E,z)\, \d{E}\\
                                                   & = & \left[\frac{\alpha}{\rm H}\right] K_{\alpha p} \frac{{\cal E_{\rm SN}}(z)}{f_b(z)\,\rho_c\,\Omega_b} f(\Lambda,x_e,\gamma,z)
\end{array}
\end{equation}
Thus, if the physical properties of the structures ($\Lambda,x_e$) as well as those of the SN ejection processes
($\gamma$) do not evolve with time, 
\begin{equation}
\left(\frac{\rm Li}{\rm H}\right)(z) =K_{\alpha p} f(\Lambda,x_e,\gamma) \int_z^\infty \frac{{\cal E_{\rm SN}}(z)}{f_b(z)\,\rho_c\,\Omega_b} \frac{\d{t}}{\d{z}}\d{z}
\label{epg}
\end{equation}
The result of the integral is shown in \Fig{upper} for both models considered here.

\subsection{Discussion}

The main parameters of this scenario are $\Lambda$ and $\gamma$.
$\Lambda$ is likely to be typically 10 g\unit{cm}{-2} in our Galaxy. 
At larger redshift, as discussed 
in \S \ref{s:igmresults}, $\Lambda$ is expected to be smaller 
due to the evolution
of both the density of the structure and of the amplitude of the magnetic field. 
A lower value of $\Lambda$ implies that CRs escape faster out of the structure, and thus produce 
less Lithium. This can be seen directly in \Eq{phism}.
Indeed, one expects a strong correlation between
$\Lambda$ and the parameter $\epsilon_{\rm shock}$  discussed earlier with
very small values of $\Lambda$ corresponding to values of $\epsilon_{\rm shock} \approx 1$.

The evolution of the abundance of \6Li can computed
 directly as a function of metallicity (see above).
 As discussed earlier, this scenario, based on the Press-Schechter
formalism will likely break down at low redshift.
If one follows the production of \6Li up to that point $z \approx 3$,
one can in fact place a limit to $\Lambda$ to avoid the over-production of \6Li
in structures.  In Model 1, we find that $\Lambda \le  10^{-5}$ g cm$^{-2}$. The upper limit
in Model 1e is a factor of 4 times larger.  These values are so low that 
our choice of $\epsilon_{\rm shock} = 1$ appears to have been well justified.
Subsequently, we expect $\Lambda$ to increase as the structure evolves into
a galaxy where standard GCR becomes important.

\section{Summary}

The observation of a \6Li plateau in halo stars at low metallicity poses
a challenge to standard models of \6Li production via Galactic
cosmic ray nucleosynthesis.  The level of the plateau
is about 1000 times larger than the standard big bang nucleosynthesis
value and about a factor of 10 larger than the GCRN value at [Fe/H] = -3.

In RVOI, we showed that an early burst of cosmic rays injected into the IGM
would produce a prompt initial enrichment of \6Li.  There, we used
the observed plateau value to normalize the energy density of CRs.
Here we applied this mechanism to a detailed model of 
cosmic chemical evolution.  The model was designed
to reproduce the observed star formation
rate at redshift $z \la 6$, the observed chemical abundances in damped Lyman alpha absorbers
and in the intergalactic medium, and to allow for an early reionization of the Universe
 at $z\sim 11$ as indicated by the third year results released by WMAP.
 As a consequence, \citet{Daigneal05}, was able to compute the 
 supernova rate as a function of redshift. This SNR was employed here
 to compute the resulting energy density in cosmic rays.
 
 Our results depend on the efficiency to which cosmic rays are
 accelerated out of the first star forming structures.  
 We found that for efficiencies, $\epsilon_{\rm shock} =1$,
 the Models 1 and 1e discussed in \citet{Daigneal05} can produce
 a plateau in \6Li at the right abundance level at low redshift when the SN energy
 deposited in CRs is about 4-15\% of the available kinetic energy. This conforms
 well with models of CR acceleration \citep{drury}. 
As in standard GCRN,  \6Li depletion at higher metallicities is necessary if the plateau is found to persist.
By computing the CR heating of the IGM, we conclude that most of the CR propagation
and hence \6Li production should be confined to the warm-hot IGM.

We also compute the in situ production of \6Li in star forming structures.
Indeed, unless CRs are allowed to escape, that is, unless 
the characteristic escape length is significantly smaller than 10 g/cm$^2$ at high
redshift the \6Li abundance would greatly exceed the observed value.  

Clearly a definitive result for the production of \6Li at high redshift or at low metallicity
will require a more detailed model for the ejection and propagation of CRs
in the early structure forming Universe.  The mechanisms described here 
would certainly produce a plateau at low metallicity.  The absolute abundance level
is uncertain but can be tied to a set of reasonable chosen physical parameters.

\acknowledgements
We are very grateful to Roger Cayrel, Tom Jones, and David Maurin for their pertinent
 and fruitful comments and Fr\'ed\'eric Daigne for 
his help. We thank E. Thi\'ebaut, and D. Munro for freely 
distributing his Yorick programming language (available at  {\em \tt
ftp://ftp-icf.llnl.gov:/pub/Yorick}), which we used to implement
part of our
analysis.  Thanks also to Pr Nath, Pr Srianand, Pr Subramanian and to 
Samui for preliminary discussions.
The work of ER was supported by a grant LAVOISIER from the French
foreign office.  The work of EV and KO has been supported by 
the collaboration INSU - CNRS France/USA. 
The work of K.A.O. was partially supported by DOE grant
DE-FG02-94ER-40823. 
\vskip 3.5cm

\end{document}